\documentstyle[12pt,world_sci]{article}
\begin{document}
\begin{flushright}
INFNCA-TH-94-18\\
August 1994
\end{flushright}

\title{{\bf EXACT SUM RULES AT FINITE TEMPERATURE AND
CHEMICAL POTENTIAL AND THEIR APPLICATION TO QCD%
\thanks{Talk presented by S.~Huang at the ``Workshop on Quantum
Infrared Physics'', Paris, France, 6--10 June 1994.
To appear in the Proceedings.}
}}
\author{
            SUZHOU HUANG\\
            {\em
   Department of Physics, FM-15, University of Washington, Seattle, \\
   Washington 98195, U.S.A.
            }\\
           and \\
           MARCELLO LISSIA\\
            {\em
   Istituto Nazionale di Fisica Nucleare, via Negri 18, Cagliari, \\
            I-09127, ITALY
            }
        }
\maketitle
\setlength{\baselineskip}{2.6ex}

\begin{center}
\parbox{13.0cm}
{ \begin{center} ABSTRACT \end{center}
  {\small
Within the framework of the operator product expansion (OPE) and
the renormalization group equation (RGE), we show that the
temperature and chemical potential dependence of the zeroth moment
of a spectral function (SF) in an asymptotically free theory is
completely determined by the one-loop structure of the theory.
This exact result constrains the qualitative shape of SF's, and
implies striking phenomenological effects near phase transitions.
  }
}
\end{center}
\vspace*{0.1cm}

  Our present understanding of QCD at finite temperature ($T$) and
baryon density (or chemical potential $\mu$) is mainly limited
in the Euclidean realm, due to the lack of non-perturbative and
systematic calculating tools directly in the Minkowski space. The
typical methods, with QCD Lagrangian as the starting point, are
the OPE and lattice simulation. Because of these two formulations
are intrinsically Euclidean, only static quantities are conveniently
studied. In order to gain dynamical informations, which are more
accessible experimentally, the analytic structure implemented
through dispersion relations often have to be invoked within the
theory of linear response.

The real-time linear response to an external source coupled to a
renormalized current $J(x)$ is given by the retarded correlator:
\begin{equation}
K(x;T,\mu)\equiv
\theta(x_0)\langle [J(x),J(0)]\rangle_{T,\mu}\, ,
\label{corr}
\end{equation}
where the average is on the grand canonical ensemble specified by
$(T,\mu)$. Disregarding possible subtraction terms, the following
dispersion relation for the frequency dependence of the retarded
correlator can be written:
\begin{equation}
\tilde{K}(\omega,{\bf k};T,\mu)=\int_{0}^\infty du^2 \,
{\rho(u,{\bf k};T,\mu)\over u^2-(\omega+i\epsilon)^2} \, .
\label{disp}
\end{equation}
For convenience, we discuss only the uniform limit (${\bf k}=0$).
Upon analytic continuation, $\omega\rightarrow iQ$, the dispersion
relation in principle connects the correlator in Euclidean region
to the SF, which embodies all the real-time information. However,
given only approximate knowledge of $\tilde{K}(iQ;T,\mu)$, either
via the OPE or lattice calculations, it is extremely difficult
technically to ``invert'' the dispersion relation. Only when we
have enough understanding of the SF involved this ``inverting''
can be made practical. In fact, the success of the so-called QCD
sum rule approach and analyzing lattice data under many
situations rests, to some extend, fortuitously on the successful
parameterization of SF's.

On the other hand, the lack of adequate understanding of SF's at
finite $(T,\mu)$ severely limits our chances to extract physical
information using these approaches. It is likely that physical
results at finite $(T,\mu)$ can be strongly biased, if one naively
assumes the same parameterizations of the SF's that work at zero
$(T,\mu)$. The purpose of this work is to derive exact sum rules
that constrain the variation of SF's with $(T,\mu)$. The derivation,
based on the OPE and the RGE, has close analogies to the analysis
of deep inelastic lepton scattering experiments. In addition, we
apply these sum rules to the chiral phase transition, and demonstrate
that SF's in some channels are drastically modified compared both
to their zero $(T,\mu)$ and perturbative shape.

In an asymptotically free theory, the OPE yields, {\em e.g.} in
the $\overline{\rm MS}$ scheme, the large-$Q^2$ asymptotic expansion
\begin{equation}
\tilde{K}(iQ;T,\mu,\kappa) \sim \tilde{K}_0(iQ,\kappa)+
\sum_n C_n(Q^2,g^2(\kappa),\kappa)
\langle [O_n]_\kappa\rangle_{T,\mu} \, ,
\label{ope}
\end{equation}
where $g^2(\kappa)$, $[O_n]_\kappa$'s and $C_n$'s are, respectively,
the coupling constant, the renormalized composite operators and their
corresponding Wilson coefficients at the subtraction mass scale
$\kappa$. It is important to notice that the information of the
ensemble average is encoded in the matrix elements of the composite
operators, while the Wilson coefficients and $\tilde{K}_0$ are
independent of $T$ and $\mu$. Although the matrix elements
$\langle [O_n]_\kappa\rangle_{T,\mu}$ cannot be determined perturbatively,
the $Q^2$-dependence of Wilson coefficients $C_n$ is dictated by the
renormalization group equation and given by
\begin{equation}
C_n(Q^2,g(\kappa),\kappa)=
{c_n\bigl(g^2(Q)\bigr)\over Q^{d_n}}\,
\biggl[{g^2(Q)\over g^2(\kappa)}
\biggr]^{(2\gamma_J-\gamma_n)/2b}
\Bigl\{1+{\cal O}(g^2(Q))\Bigr\}\, ,
\label{cn}
\end{equation}
where $d_n$ is the canonical dimension of the operator $O_n$ minus
the dimension of $\tilde{K}$ and $c_n(g^2(Q))$ is calculable
perturbatively. The pure numbers $\gamma_i$ ($i=J$, $n$) and $b$
are related to the anomalous dimensions of $J$, $O_n$ and to the
$\beta$-function as follows
\begin{equation}
\Gamma_i=-\gamma_i g^2+{\cal O}(g^4)\, ,\,\,\, {\rm and}
\,\,\, \beta=-b g^4+{\cal O}(g^6) \, .
\label{rgef}
\end{equation}

  To study the dependence of $\tilde{K}$ on $(T,\mu)$ we only need
to consider the difference
$
\Delta\tilde{K}(iQ) \equiv \tilde{K}(iQ;T,\mu)-\tilde{K}(iQ;T',\mu')
$
and
\begin{equation}
\Delta\tilde{K}(iQ)
=\int_0^\infty du^2 {\Delta\rho(u)\over u^2+Q^2}\, ,
\label{ddisp}
\end{equation}
where $\Delta\rho(u)\equiv\rho(u;T,\mu)-\rho(u;T',\mu')$.
This subtraction is crucial to remove $\tilde{K}_0(iQ,\kappa)$,
which contains all the terms not suppressed by a power of $1/Q^2$,
and also to make $\Delta\tilde{K}(iQ)$ independent of the
renormalization point $\kappa$. Finite masses give corrections of
order $m^2(Q)/Q^2$, with $m^2(Q)$ that runs logarithmically and
hence can be ignored, if we are only interested in the lowest moment
of the subtracted SF.

At this point we have expressed the left-hand side of
Eq.~(\ref{ddisp}) as an asymptotic expansion of the form:
\begin{equation}
\Delta\tilde{K}(iQ)\sim
\sum_{n,\nu = 0}^{\infty}
\frac{c_n^{(\nu)}(\kappa)\Delta\langle [O_n]_\kappa\rangle}{Q^{d_n}}
[g^2(Q)]^{\nu+\eta_n}\, ,
\label{asyope}
\end{equation}
where $\Delta\langle[O_n]_\kappa\rangle$ denotes the difference
between the expectation values of $[O_n]_\kappa$ in the ensembles
specified by $(T,\mu)$ and $(T',\mu')$ and the exponent $\eta_n$
and the $Q^2$-independent coefficients $c_n^{(\nu)}(\kappa)$ are
known perturbatively.

We proceed by making an analogous asymptotic expansion of
$\Delta\rho(u)$:
\begin{equation}
\Delta\rho(u)\sim
 \sum_{n=0}^\infty
\frac{[g^2(u)]^{\xi_n}}{u^{2(n+1)}}\,
\sum_{\nu=0}^\infty
a_n^{(\nu)}[g^2(u)]^{\nu}\, .
\label{rholn}
\end{equation}
For notational clarity, we have ignored exponentially suppressed terms
and the fact that there can be more than one $\eta_n$ and $\xi_n$ for
each $n$. We then obtain the sum rules by imposing that the asymptotic
expansion of the right-hand side of Eq.~(\ref{ddisp}), which we get by
inserting Eq.~(\ref{rholn}) in the dispersion integral, matches the
left-hand side obtained by the OPE. In a long paper~\cite{longpaper} we
shall present those technical details that make the procedure sketched
above rigorous. At the moment we only need to consider the leading terms
in the expansions of, respectively, the left-hand side and the right-hand
side of Eq.~(\ref{ddisp}):
\begin{equation}
[g^2(Q)]^{\eta_n}
\frac{\Delta\langle [O_n]_\kappa\rangle}{Q^{d_n}}
[c_n^{(0)}(\kappa) + c_n^{(1)}(\kappa)g^2(Q)]\, ,
\label{leadinglh}
\end{equation}
\begin{equation}
{1\over Q^2}\biggl[\overline{\Delta\rho}\,
+{a_{0}^{(0)}\over(1-\xi_n)}[g^2(Q)]^{\xi_n-1}\biggr]\, .
\label{leadingrh}
\end{equation}
In Eq.~(\ref{leadingrh}), if $\xi_n>1$, $\overline{\Delta\rho}$
can be shown to be equal to the zeroth moment of the subtracted SF;
note that the zeroth moment of a function whose
asymptotic expansion is Eq.~(\ref{rholn}) is infinite, if $\xi_n\leq 1$,
since $\int_A^\infty \! dx\, x^{-1}(\log x)^{-\xi_n} = \infty$.

For the sake of concreteness, let us examine the consequences
of matching Eq.~(\ref{leadinglh}) and Eq.~(\ref{leadingrh})
in the case we are concerned with, {\em i.e.} $n=1$ and $d_1=2$.
If the OPE calculation produces $\eta_n=0$, then $\xi_n$ must be an
integer greater than one, the zeroth moment exists and is given by
\begin{equation}
\int_0^\infty \!\!\! du^2\, \Delta\rho(u) =
c_1^{(0)}(\kappa)\Delta\langle [O_1]_\kappa\rangle\, .
\label{cons}
\end{equation}
If $\eta_n>0$, then $\xi_n=1+\eta_n$, the zeroth moment is again
finite and equal to zero:
\begin{equation}
\int_0^\infty \!\!\! du^2\, \Delta\rho(u) = 0\, .
\label{nonc}
\end{equation}

Our main results, Eq.~(\ref{cons}) and Eq.~(\ref{nonc}), can be
expressed in physical terms as follows. The zeroth moment of a SF for
a current $J$ whose OPE expansion yields $\eta_n > 0$ is independent
of $T$ and $\mu$, while the same moment for a current with $\eta_n=0$
changes with $T$ and $\mu$ proportionally to the corresponding
change(s) of the condensate(s) of leading  operator(s). When $\xi_n<1$
or $\xi_n=1$ in Eq.~(\ref{rholn}) terms such as $[g^2(Q)]^{-1}$ or
$\ln[g^2(Q)]$ in Eq.~(\ref{leadingrh}) would be produced. Hence the
appearance of the inverse power of the $g^2(Q)$ or $\ln[g^2(Q)]$ in
the OPE series is a indication that the zeroth moment of the subtracted
SF is infinite.

At this point several more general comments are appropriate: 1) It is
essential to take into account the QCD logarithmic corrections, for
the logarithmic corrections not only dictate whether
$\overline{\Delta\rho}$ satisfy Eq.(\ref{cons}) or Eq.(\ref{nonc}),
but also control the very existence of $\overline{\Delta\rho}$. 2)
The derivation of sum rules for higher moments of the SF requires the
complete cancelation of all the lower dimensional operator terms, not
just the leading $g^2(Q)$ terms; in particular, we also need current
quark mass corrections to the Wilson coefficients. An appropriate
subtraction is the prerequisite for the convergence of higher moments.
3) The $(T,\mu)$-dependent part of the leading condensate appearing
in Eq.~(\ref{cons}) does not suffer from the infrared renormalon
ambiguity, because only the perturbative term $\tilde{K}_0$ can
generate contributions to the leading condensate that are dependent
on the prescription used to regularize these renormalons. But
$\tilde{K}_0$ is independent of $T$ and $\mu$ and any prescription
dependence cancels out when we make the subtraction in Eq.~(\ref{ddisp}).
On the contrary, unless we generalize Eq.~(\ref{ddisp}) and make other
subtractions, sum rules that involve non-leading condensates are, in
principle, ambiguous. 4) We have explicitly verified the correctness of
our results in a soluble model~\cite{longpaper}, the Gross-Neveu model
in the large-$N$ limit, where we obtain exactly all the relevant
quantities, such as SF's at arbitrary $(T,\mu)$, Wilson coefficients,
$\beta$- and $\Gamma$-functions in pseudoscalar and vector channels.

  Now let us specialize to QCD and consider four correlation functions:
two involving the non-conserved scalar $J_S=\bar{\psi}\psi$ and
pseudoscalar $J_P=\bar{\psi}\gamma_5\psi$ currents
($\gamma_{J_S}=\gamma_{J_P}=1/4\pi^2$) and two involving the conserved
vector $J_V=\bar{\psi}\gamma_\mu\psi$ and axial-vector
$J_A=\bar{\psi}\gamma_\mu\gamma_5\psi$ currents ($\gamma_V=\gamma_A=0$).
Since the leading operators (dimension four) have non-positive anomalous
dimensions, the two non-conserved currents have
$\eta_n\geq (2\gamma_J-\gamma_n)/2b>0$ and Eq.~(\ref{nonc}) applies,
{\em i.e.} the zeroth moments of their SF's are independent of $T$ and
$\mu$. On the other hand, the two conserved currents have $\eta_n=0$
and a generalization of Eq.~(\ref{cons}) applies~\cite{longpaper}.
Since there are three dimension-four operators with zero anomalous
dimension,
the sum rules for the vector and axial-vector currents are
\begin{equation}
\int_0^\infty\!\! du^2\, \Delta\rho(u)=
a\Delta\langle[m\bar{\psi}\psi]\rangle
+{\Delta\langle[\alpha_s G^2]\rangle\over 2\pi}
+8\Delta\langle [\theta_{00}]\rangle\, ,
\label{va}
\end{equation}
where $\theta_{\mu\nu}$ is the traceless stress tensor and $a=6$ for
vector and $a=-10$ for axial channels respectively. This exact sum
rule should not be contaminated explicitly by instantons, although
the value of the condensates certainly have instanton contributions.
The reason is that the instanton singularities in the Borel-plane
are located on the positive axis starting at $8\pi^2$, and, therefore,
contribute to correlation functions only with higher order terms of
$1/Q^2$ in the OPE series.

Finally, let us discuss some of the phenomenological consequences
of these exact sum rules. In the pseudoscalar channel,
$\overline{\Delta\rho}=0$ implies that, in the broken-chiral-symmetry
phase, the change of the pion pole induced by $T$ or $\mu$
is exactly compensated by a corresponding change of the continuum
part of the SF. Next let us consider the scalar correlation
function at $Q^2=0$, the chiral susceptibility,
\begin{equation}
\chi(T,\mu)\equiv
\int\! d^4x\, \theta(x_0)\langle [J_S(x),J_S(0)]\rangle_{T,\mu}
=\int_0^\infty\!\! du^2\, {\rho(u;T,\mu)\over u^2}\, ,
\label{chirals}
\end{equation}
which diverges when $(T,\mu)$ approaches the phase boundary, provided
the chiral restoration is a continuous transition. The divergence of
the chiral susceptibility near phase transition only can be produced
in Eq.(\ref{chirals}) by singularities very close to the origin, when
the exact sum rule $\overline{\Delta\rho}=0$ is simultaneously taken
into account. Thus the spectral function would have to have a vanishing
threshold (since there is no massless poles in the chirally symmetric
phase) and develop a strong peak right above the threshold, when $(T,\mu)$
is very close to the phase boundary. Because in the chirally symmetric
phase the pseudoscalar and scalar channels are degenerate, the same would
also happen to the pseudoscalar SF. This strong peak in the pseudoscalar
and scalar SF's, which is intimately connected with the critical
phenomenon of diverging susceptibility and the correlation length near
the phase transition, can be interpreted as some kind of quasi-particle,
thus confirming the qualitative picture, originally proposed in the
context of the Nambu-Jona-Lasinio model~\cite{kunihiro}, of the
appearance of soft modes near the chiral phase transition. Similar
situation, though less drastic, can be argued to occur in vector and
axial channels, based on the facts that
$\Delta\langle [\theta_{00}]\rangle$,
$\Delta\langle [m\bar{\psi}\psi]\rangle$ and
$\Delta\langle[\alpha_s G^2]\rangle$ behave smoothly across
the critical line and the rapid increase of the so-called baryon
number susceptibility at the chiral restoration point in lattice
simulations~\cite{gottlieb}.
If the chiral restoration turns out not to be a second order phase
transition, but rather a cross-over or weak first order transition
(finite but large correlation length), as the lattice data seem to
indicate~\cite{phase}, we expect the same qualitative features,
though less pronounced.

  In summary, we used OPE and RGE to derive exact sum rules at finite
$(T,\mu)$ in asymptotically free theories. These exact sum rules
strongly constrain the qualitative shape of SF's, especially near
phase transition region. We urge whoever parameterizes a SF, {\em e.g.}
in the QCD sum rule calculations or to interpret lattice simulations,
to incorporate these exact constraints. In the future, we plan to
generalize these results to baryonic currents and analyze their
phenomenological consequences in greater detail.

  We thank Profs. Fried and M\"uller for organizing a very enjoyable
workshop. This work is supported in part by the US Department of Energy.

\vspace*{0.1cm}
\bibliographystyle{unsrt}

\begin{thebibliography}{99.}
\bibitem{longpaper} S.~Huang and M.~Lissia, hep-ph/9404275, and
a manuscript in preparation.

\bibitem{kunihiro} T.~Hatsuda and T.~Kinihiro, Phys. Rev. Lett.
{\bf 55}, (1985) 158.

\bibitem{gottlieb} S.~Gottlieb et al., Phys. Rev. {\bf D38}, (1988) 2888.

\bibitem{phase} F.~Brown et al., Phys. Rev. Lett. {\bf 65}, (1990) 2491.

\end{thebibliography}

\end{document}